\documentclass[12pt, prd, showpacs,superscriptaddress,notitlepage]{revtex4-1}

\usepackage[a4paper, margin=0.8in]{geometry}
\usepackage{setspace}

\usepackage{amssymb}
\usepackage{amsmath}

\usepackage{graphicx}

\usepackage{color}
\definecolor{darkblue}{rgb}{0,0,1}
\definecolor{darkgreen}{rgb}{0,.4,0}
\definecolor{deepred}{rgb}{.5,0,0}
\usepackage[colorlinks,linkcolor=darkblue,urlcolor=deepred,citecolor=darkgreen,unicode]{hyperref}

\begin{document}
\title{A new ripplon branch in He~II}
\author{I. V. Tanatarov}
\email{igor.tanatarov@gmail.com}
\affiliation{National Science Center ``Kharkov Institute of Physics and Technology'',
 Academicheskaya St. 1, Kharkov, 61108, Ukraine.}
\author{I. N. Adamenko}
\email{i.n.adamenko@mail.ru}
\author{K.E. Nemchenko}
\affiliation{Karazin Kharkov National University,
        Svobody Sq. 4, Kharkov, 61077, Ukraine.}
\author{A.F.G.~Wyatt}
\affiliation{School of Physics, University of Exeter,
          Exeter EX4 4QL, UK.}

\begin{abstract}
We analyse the dispersion relation of ripplons, on the surface of superfluid helium, using the dispersive hydrodynamics approach and find a new ripplon branch. We obtain analytical equation for the dispersion relation and  analytic expressions for the limiting cases. The probabilities of decay of unstable ripplons above the roton gap into rotons are derived. A numerical solution for the ripplon dispersion curve is obtained. The new ripplon branch is found at energies just below the instability point of the bulk spectrum, and is investigated; its stability is discussed. 
\end{abstract}

\maketitle

\section{Introduction}

\begin{figure}[!ht]
\begin{center}
\includegraphics[viewport=77 254 463 522, width=0.75\textwidth]{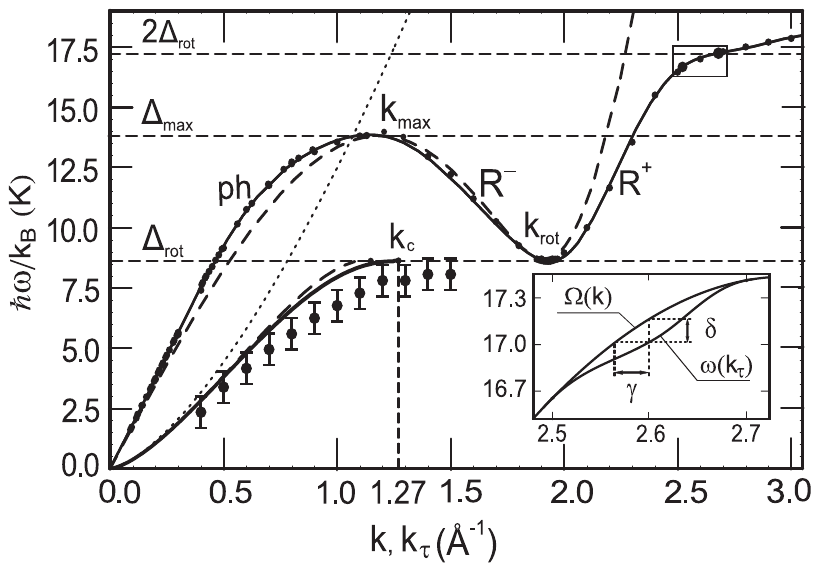}
\parbox{0.9\textwidth}{\caption{\label{Fig1} The dots are experimental data for the dispersion relation of bulk excitations $\Omega(k)$ \cite{neutron}, and the thin line shows its analytic approximation that is used, with $S\!=\!18$. The thick line shows the results of numerical solution of Eq. (\ref{RipplonDispersionNum}) for the ripplons' dispersion $\omega(k_{\tau})$. The dashed lines are the approximations of $\Omega(k)$ used in \cite{Ripplon} and the resulting curve $\omega(k_{\tau})$ obtained there. Dotted line shows the $\sim k^{3/2}$ law. Large dots with error bars are experimental data for the ripplon dispersion \cite{exp}. Two black dots in the high-energy part of the bulk spectrum designate the end points of the very high energy ripplon solution, and the insert graph shows its behaviour schematically, where $\gamma\!=\!0.7\cdot 10^{-3}${\AA}${}^{-1}$ and $\delta\!=\!1.6\cdot 10^{-3}K$.}}
\end{center}
\end{figure}

Superfluid helium is a unique medium, in which there are well defined long-living excitations with very short wavelengths, of the order of average interatomic distance, -- rotons. The continuous dispersion curve of the bulk excitations contains both the long-wavelength phonon part, which is almost linear, and the essentially nonlinear maxon-roton part (see Fig.~\ref{Fig1}).

Ripplons are quantised capillary waves on the free surface of superfluid $^4$He. At low frequencies their dispersion law gives a good way of measuring the surface tension of liquid helium \cite{KingWyatt,Rochesurften}.  The temperature dependence of the surface tension is due to ripplons \cite{Atkins-ripplons}. Ripplons can be  detected by neutron scattering, in a similar way to the bulk modes, and  have been shown to exist up to wavenumber of 1.5 $\textrm{\AA} ^{-1}$, where the ripplon energy is close to that of  the roton minimum, 8.6 K, \cite{exp,exp2}. Ripplons are the dominant scatterer of surface state electrons on liquid helium (see \cite{GrimesAdams,DeboraCoimbra} and references therein). Ripplons play a significant role in the condensation \cite{brownwyatt}, evaporation and reflection of atoms from liquid $^4$He, \cite{NayakEdwards,wyatttuckercregan}. It has been suggested  that ripplons are the most favourable excitations for the simulation of general-relativistic effects related to horizons of white holes \cite{volovik}.

It was shown \cite{exp,exp2}, that the surface modes of He~II, ripplons, exhibit the same properties as the bulk modes, being well-defined quasiparticles even in the high-energy region, close to the roton gap. Their dispersion relation in this region is determined by the bulk excitations, in particular rotons. It was studied in theory in work \cite{P&S92}, and in \cite{P&S95} the ripplon dispersion curve below the roton gap was obtained in the framework of density functional approach, and it was in good agreement with the experimental data.

In this paper we present  a theoretical model of ripplons which is good enough to account for the measured dispersion curve $\omega(k)$ but is simple enough to expose the underlying physics. It explains why ripplons only exist in certain parts  of the ($\omega,k$) plane, and why the  dispersion curve approaches the line  $\omega=\Delta_{rot}$ at the top of a parabola, where $\Delta_{rot}$ is the energy of the roton minimum. Moreover the model predicts a new roton branch at energies $< \sim 2\Delta_{rot}$.

The theoretical model is a non-local hydrodynamical theory \cite{P&S92}, developed by us in Ref. \cite{PRB}. The actual physical characteristics of the liquid are introduced into the model by using the measured dispersion curve for bulk excitations. The justification of application of nonlocal hydrodynamics to the description of a quantum fluid at interatomic distances is given in \cite{PRB}, and in more detail in \cite{PRB2008}. This model was used in \cite{PhNT}, \cite{JLTP2006} and \cite{PRB2008} to describe interaction of He~II phonons and rotons with solid interfaces. It is based on the fact, that in a quantum fluid the atoms are delocalized, as their thermal de Broglie wavelength is greater than the atom spacing. The general idea of description of a quantum fluid in terms of hydrodynamic variables at interatomic scales is being widely used. One of the first to exploit it was Atkins in 1959 \cite{Atkins-hydro}, when he introduced bubbles and snowballs of microscopic size in order to describe the mobility of electrons and ions in He II by means of methods of theory of continuous medium. The idea is utilized now in the variations of density functional theory (see for example \cite{P&S95}), and is implicitly assumed in other fields of research related to superfluidity. In particular, the variables of continuous media are used lately for description of vortices in helium with dimensions of the order of the interatomic distances (see for example \cite{Natsik}).

The quantum fluid, which is considered continuous at any length scales, can be described in terms of the variables of continuous media, i.e. density, pressure and velocity, which satisfy the mass and momentum conservation laws. The equations of ideal liquid, which follow from them, do not form a closed system, and are supplemented by the equation of state, which specifies the functional relation between pressure and density. As shown in \cite{PRB,PRB2008}, the equation of state for small deviations of the system from equilibrium in the general case is non-local, with some difference kernel $h(\mathbf{r})$.

The closed linear system of equations is brought to a non-local integro-differential wave equation with regard to pressure. The dispersion relation of the fluid $\Omega(k)$ then is determined by the Fourier transform of the kernel $h(\mathbf{r})$ and can contain arbitrary degrees of $k$. In this paper we start from the dispersion relation that approximates the experimental data for the spectrum of superfluid helium. The chosen $\Omega(k)$ gives explicit expression for the kernel $h(\mathbf{r})$, which determines the equation of state and the non-local wave equation.

In order to derive the dispersion relation of ripplons, we look for the solution of the nonlocal wave equation (\ref{EQP}) in half-space, equipped with usual boundary conditions (\ref{boundary-landau}). The use of simple model of superfluid helium allows us to fully solve the problem of ripplons' spectrum. The ripplons' dispersion equation is obtained in algebraic form, and its analytic solutions are derived in limiting cases. The analytic and numerical results of the paper in the region below the roton gap are consistent with the experiments, numerical computations and qualitative estimates of other authors \cite{exp,P&S95}.

A preliminary and partial account of this analysis is given in \cite{Ripplon}, where a very approximate function for the dispersion curve for bulk phonons and rotons was used. This allowed to derive in \cite{Ripplon} the equation for the ripplon's dispersion relation in elementary functions. Its solution below the roton gap yielded good agreement with the experiment and previous works. However, the simplicity of the used function $\Omega(k)$ also limited the obtained results to semi-quantitative and restricted the investigation of surface excitations to the energies below the roton gap.

In this work we follow the approach used in \cite{Ripplon}, but now we use a much more detailed function $\Omega(k)$, which gives approximation of the experimentally measured curve in the whole interval of wave vectors with good accuracy. In fact, the proposed scheme works for functions $\Omega(k)$ that approximate the experimental data with any given precision. This allows us to obtain accurate results in the whole energy interval up to twice the roton gap energy, which is the Pitaevskii instability point of the bulk excitations spectrum \cite{EndSpectrum}, and in particular to search for the surface solutions above roton gap, with wave vectors greater than the $R^+$ rotons'. In the second section we formulate the mathematical problem. It is brought to the parametric equation for the ripplons' dispersion relation in the next section. The equation is investigated both at small energies and high energies, close to the roton gap and up to the instability point. It is just below this energy where we find the new unusual ripplon branch. The numerical solution is given and discussed in the fourth section of this work, the new branch is investigated and its stability is discussed.

\section{Equations and boundary conditions}

Let us consider the half-space $z\!>\!0$ filled by superfluid helium. In accordance with the approach of \cite{PRB}, it obeys the ordinary linearized equations of an ideal liquid, but the relation between the deviations $P$ and $\rho$ of pressure and density from the respective equilibrium values is nonlocal, with some difference kernel $h(\mathbf{r})$. The problem in terms of pressure can be expressed as a nonlocal wave equation. When solving the problem in half-space, the integration domain is limited to this half-space \cite{PRB2008} and the problem can be brought to the form
\begin{equation}
\label{EQP}
\triangle
P(\mathbf{r},t)=
\!\int\limits_{z_{1}>0}\! d^{3}r_1\,
    h(|\mathbf{r}-\mathbf{r}_1|)\ddot{P}(\mathbf{r}_1,t), \quad
        x,y,t\in(-\infty,\infty), \, z\in(0,\infty).
\end{equation}
We assume that the interface is sharp enough to consider that the kernel $h(r)$ is the same in the presence of the interface as it is in the bulk medium. We discuss possible consequences of taking into account the smoothness of the density profile of the free surface at the end of section 4. The kernel is related to the dispersion relation of the bulk excitations $\Omega(k)$ through its Fourier transform (see \cite{PRB})
\begin{equation}
        h(k)=\frac{k^{2}}{\Omega ^{2}(k)}.
\end{equation}
Thus the equation describes a continuous medium with dispersion relation $\Omega(k)$, which fills the half-space $z\!>\!0$. The relation $\Omega(k)$ can be essentially nonlinear and have the distinctive form of the dispersion of superfluid helium. The latter is almost linear at small wave vectors, reaches the maxon maximum $\Delta_{max}$ at wave vector $k_{max}$, then after the decreasing part reaches the roton minimum $\Delta_{rot}$ at $k_{rot}$, and increases again until the instability point at frequency $2\Delta_{rot}$ \cite{EndSpectrum} (see Fig.~\ref{Fig1}).

It is important, that the problem (\ref{EQP}) is stated in the half-space $z\!\in\!(0,\infty)$, and cannot be solved by an even continuation of the solution to negative $z$ with extension of integration limits to infinity, as was done in \cite{P&S92}. Such continuation gives a mathematical problem, which is completely different from (\ref{EQP}) and is not physically consistent, in the general case violating energy conservation at the interface (see more in \cite{PhNT}). Indeed, if Eq. (\ref{EQP}) was true on $z\!\in\!(-\infty,\infty)$, then the values of $P$ at $z\!<\!0$ would be determined by the values of $P$ at $z\!>\!0$, and the resulting solution would not be even.

The problem (\ref{EQP}) in one dimension was solved in \cite{PhNT} for arbitrary dispersion with the help of Wiener and Hopf method, in the case where only one root is real, as it is for energies less than $\hbar\Delta_{rot}$. In \cite{JLTP2006} the solution was generalized to 3D, and in \cite{PRB2008} it was generalized to the case when the function $\Omega^{2}(k^2)$ is a third degree nonmonotonic polynomial. This is the simplest analytic form of $\Omega^{2}(k^2)$, which can qualitatively reproduce the distinctive dispersion of superfluid helium, including both the phonon and the maxon-roton parts. In \cite{Ripplon} the same solution was used for a study of the dispersion relation of ripplons. The use of such simple approximation of $\Omega^{2}(k^2)$ allowed to derive the equation for the ripplon dispersion relation in elementary functions, but also limited the obtained results to semi-quantitative. In this work we investigate the dispersion relation of ripplons, basing on this equation, but for the case when $\Omega^{2}(k^2)$ is a polynomial of high power, which can approximate the experimentally measured curve with any needed precision.

The equation (\ref{EQP}) is supplemented by the boundary condition on the free surface. The pressure at the surface with surface tension $\sigma$ is the Laplace pressure $P_{L}\!=\!\sigma/R$, where $R$ is the surface curvature radius. For small deviations of the surface from equilibrium position this can be rewritten as
\begin{equation}
    \label{boundary-landau}
    P=\sigma \left(\frac{d^{2}\xi}{dx^{2}}+\frac{d^{2}\xi}{dy^{2}}\right),
\end{equation}
where $\xi$ is the $z$-coordinate of the points of the surface (see for example \cite{P&S92} or \cite{L&L}). For a solution $\sim\exp(i\mathbf{kr}\!-\!i\omega t)$, with given frequency $\omega$ and the projection $\mathbf{k}_{\tau}$ of wave vector on the plane of the free surface $(x,y)$, this expression turns into $P\!=\!-\sigma k_{\tau}^{2}\xi$. The $z$-component of velocity of the surface in this case is $\mathrm{v}_{z}\!=\!-i\omega\xi$, and therefore we can write the boundary condition in the form
\begin{equation}
    \label{boundary}
   \left. \mathrm{v}_{z}\right|_{z=0}=
    \frac{i\omega}{\sigma k_{\tau}^{2}}
    \left.P\right|_{z=0}.
\end{equation}

Substitution of the solution of Eq. (\ref{EQP}) into this boundary condition will give us a parametric equation for the dispersion relation of ripplons $\omega(k_{\tau})$.

It should be noted, that in reality the density profile of the free surface of superfluid helium is rather smooth, with a transition layer of several monolayers thickness. We discuss possible consequences of taking this into account at the end of section 4.

\section{Equation for the ripplons' dispersion and its analytic solution}

Let us assume now that the function $\Omega^{2}(k^2)$ is a polynomial of degree $S$ in powers of $k^{2}$, such that the only real zero of $\Omega^{2}(k^2)$ is $k^{2}\!=\!0$, where $\Omega^{2}\sim k^2$. It was shown in \cite{JLTP2006}, that in this case the Fourier image in terms of $\mathbf{r}$ and $t$ of the solution of equation (\ref{EQP}) of the simplest form is
\begin{equation}
    \label{Solution}
        P(k_z;\omega,\mathbf{k}_{\tau})=
        \frac{C_{out}(\omega,\mathbf{k}_{\tau})}{k_{z}-k_{1\,z}(\omega,\mathbf{k}_{\tau})}
        {\prod\limits_{k_{iz}\in\mathcal{C}_{+}}}\!\!\!'
        \frac{k_{z}-k_{i\,z}(\omega\!=\!0,\mathbf{k}_{\tau})}{k_{z}-k_{i\,z}(\omega,\mathbf{k}_{\tau})}
\end{equation}
Here $\mathbf{k}\!=\!\mathbf{k}_{\tau}\!+\!\mathbf{e}_{z}k_{z}$ is the wave vector. The product is taken over all the roots $k_{z}\!=\!k_{i\,z}$ of equation
$\Omega^{2}(k^{2}\!=\!k_{\tau}^{2}+k_{z}^{2})\!=\!\omega^2$ in the upper half-plane $\mathcal{C}_{+}$ of the complex variable $k_z$. The real roots are
assumed to be shifted from the real line in accordance to some selection rules (see \cite{PhNT}). Different selection rules regarding to which roots are shifted up and which down lead to different linear independent solutions. There is a condition that the number of roots shifted up and down should be equal, so the full number of roots $k_{i\,z}$ in $\mathcal{C}_{+}$ is $S$. The root $k_{1\,z}(\omega,\mathbf{k}_{\tau})$ is the phonon root, i.e. the one in $\mathcal{C}_{+}$, which continuously turns to zero at $\omega\!=\!0$ and $k_{\tau}\!=\!0$. The prime superscript on the product designates that $k_{1\,z}$ is omitted in it. Function $C_{out}$ is the amplitude.

Taking the inverse Fourier transform, we obtain $P(\mathbf{r},t)$, and then velocity of continuous medium $\mathbf{v}(\mathbf{r},t)$ is found from the relation $\dot{\mathbf{v}}\!=\!-\nabla P/\rho_{0}$. The solutions with given $\omega$ and $\mathbf{k}_{\tau}$ are
\begin{equation}
    \label{solution-waves}
    P(\mathbf{r},t;\mathbf{k}_{\tau},\omega)\!=\!\sum\limits_{k_{iz}\in\mathcal{C}_{+}}\!r_{i}e^{i\mathbf{k}_{i}\mathbf{r}-i\omega t};\quad
    \mathbf{v}(\mathbf{r},t;\mathbf{k}_{\tau},\omega)\!=\!\frac{1}{\rho_{0}\omega}\sum\limits_{k_{iz}\in\mathcal{C}_{+}}\!\mathbf{k}_{i}r_{i}e^{i\mathbf{k}_{i}\mathbf{r}-i\omega t}.
\end{equation}
Here $\mathbf{k}_{i}\!=\!\mathbf{k}_{\tau}+\mathbf{e}_{z}k_{i\,z}$, the sums are taken over the same roots $k_{i\,z}$ as in (\ref{Solution}) but including $k_{1\,z}$; $r_{i}$ is the residue of the right hand part of (\ref{Solution}) in $k_{i\,z}$.

In works \cite{PhNT}, \cite{PRB2008} and \cite{JLTP2006}, where the prime concern was the problem of interaction of waves with the interface between helium and a solid, the solutions which contained at least one running wave were sought. So there were from one to three real roots $k_{i\,z}$, and different rules of shifting them from the real line, that determine the set of real $k_{i\,z}$ in (\ref{Solution}) and their signs, were used to obtain different linear independent solutions, of which the general solution was composed.

In the current work we are interested in ripplons, and therefore in the surface solutions that are damped away from the interface. So we have to take that all the $k_{i\,z}$, for all $i$, are not real. Then the selection rules are no longer needed, and the surface solution is unique. It still has the form (\ref{Solution}), but now we investigate it in a different region on the plane of parameters $(\omega,k_{\tau})$, in which $k_{\tau}^{2}$ is greater than any of the real roots $k_{i}^{2}(\omega)$ of equation $\Omega^{2}(k^{2})=\omega^{2}$.

We are interested in the dispersion relation that approximates the one of superfluid helium (see Fig.~\ref{Fig1}). Then the mentioned region on the plane $(\omega,k_{\tau})$ can be divided into two parts. The first is below the roton minimum $\omega\!<\!\Delta_{rot}$ and on the low $k$ side of the phonon dispersion curve $k_{\tau}\!>\!k_{1}$. The second is above the roton minimum $\omega\!>\!\Delta_{rot}$ and on the high $k$ side of the $R^{+}$ roton curve $k_{\tau}\!>\!k_{3}$. Here it is supposed that the subscripts are assigned to the real roots $k_{i}(\omega)$ of equation $\Omega^{2}(k)\!=\!\omega^{2}$ in the ascending order of their absolute values: there are only phonons $k_{1}$ below $\Delta_{rot}$, and at $\omega\!>\!\Delta_{rot}$ we have $k_{1}\!<\!k_{2}\!<\!k_{3}$, so $k_{2}$ corresponds to $R^-$ rotons and $k_{3}$ to $R^{+}$ rotons.

The values of $P$ and $\mathrm{v}_{z}$ on the surface $z\!=\!0$ are obtained from (\ref{solution-waves}) as the sums of the written residues in all the finite singular points of the corresponding expressions, because the latter are built so that all of their singularities lie in $\mathcal{C}_+$. When calculating $\mathrm{v}_{z}$, we take into account that $k_{i\,z}\mathrm{res}_{k_{z}\rightarrow k_{i\,z}}P(k_{z})\!=\!\mathrm{res}_{k_{z}\rightarrow k_{i\,z}}\big(k_{z}P(k_{z})\big)$. Then the values $\left.P\right|_{z=0}$ and $\left.\mathrm{v}_{z}\right|_{z=0}$ are equal to minus the residues of $P(k_{z})$ and $k_{z}P(k_{z})/(\rho_{0}\omega)$ in infinity correspondingly, and those are obtained directly by expanding $P$ from (\ref{Solution}). Then it can be shown that
\begin{equation}
    \label{Solution-surface}
    \left.\mathrm{v}_{z}\right|_{z=0}=\frac{1}{\rho_{0}\omega}
    \bigg\{k_{1\,z}+\sum\limits_{i=2}^{S}\big[k_{i\,z}-k_{i\,z}(\omega\!=\!0)\big]\bigg\}
    \left.P\right|_{z=0}.
\end{equation}

Comparing (\ref{Solution-surface}) with the boundary condition (\ref{boundary}), we finally obtain the equation for the dispersion relation $\omega(k_{\tau})$ of ripplons:
\begin{equation}
    \label{RipplonDispersion}
    \omega^2=\frac{\sigma}{\rho_{0}}\cdot k_{\tau}^2\frac{1}{i}
    \bigg\{k_{1\,z}(\omega,k_{\tau})+
	\sum\limits_{i=2}^{S}\big[k_{i\,z}(\omega,k_{\tau})-k_{i\,z}(\omega\!=\!0,k_{\tau})\big]\bigg\}.
\end{equation}
This equation is written in terms of $k_{i\,z}(\omega,k_{\tau})=\sqrt{k_{i}^{2}(\omega)-k_{\tau}^{2}}\!\in\!\mathcal{C}_+$, where $k_{i}^{2}(\omega)$ in turn are the roots of the polynomial equation $\Omega^{2}(k^{2})\!=\!\omega^2$ with real coefficients. Then taking into account the condition that all the $k_{i\,z}$ are non-real, we can show, that the right hand part of the equation (\ref{RipplonDispersion}) is in fact real. Indeed, let us consider the equation $\Omega^{2}(k^{2}\!=\!k_{z}^{2}\!+\!k_{\tau}^{2})\!=\!\omega^2$ with regard to $k_{z}^{2}$. This equation is also a polynomial equation with real coefficients. So its roots with regard to $k_{z}^{2}$ are either negative (no positive roots exist or there would be real $k_{i\,z}$) or break up into complex-conjugate pairs. The negative roots give imaginary $k_{i\,z}$ and complex-conjugate pairs $k_{i\,z}^{2},k_{j\,z}^{2}$ give pairs $k_{i\,z},k_{j\,z}$ in $\mathcal{C}_{+}$, such that $k_{i\,z}\!=\!-k_{j\,z}^{\ast}$, and for such a pair $k_{i\,z}+k_{j\,z}\!=\!i(\mathrm{Im}\, k_{i\,z}\!+\mathrm{Im}\,k_{j\,z})\!=\!2i\mathrm{Im}\,k_{i\,z}$. Therefore the expression inside the braces in (\ref{RipplonDispersion}) is imaginary and the right hand part is real.

In the limit of small frequencies $\omega\!\rightarrow\!0$ the sum in (\ref{RipplonDispersion}) tends to zero and the equation turns into
\begin{equation}
    \omega^{2}\!=\!\frac{\sigma}{\rho_{0}}\cdot k_{\tau}^{2}\sqrt{k_{\tau}^{2}-\frac{\omega^{2}}{s^{2}}},
\end{equation}
where $s$ is sound velocity at zero frequency. This is also the form that the equation takes in the case when the dispersion relation of the fluid is linear. Its exact solution is
\begin{equation}
    \label{compressibility}
    \omega^{2}_{lin}=\frac{\sigma}{\rho_{0}}\cdot k_{\tau}^{3}
    \left\{\sqrt{1+\Big(\frac{\sigma}{2\rho_{0}s^2}k_{\tau}\Big)^2}-\frac{\sigma}{2\rho_{0}s^2}k_{\tau}\right\}.
\end{equation}
The primary term, which is left if we neglect compressibility of the fluid (i.e. in the limit $s\!\rightarrow\!\infty$), gives the relation, well-known from hydrodynamics \cite{L&L}
\begin{equation}
    \label{three-half}
    \omega^{2}=\frac{\sigma}{\rho_{0}}\cdot k_{\tau}^{3}.
\end{equation}

In order to obtain further expansion of $\omega$ in powers of $k_{\tau}$, we take into account that for small frequencies, accurately within $O(\omega^4)$, $k_{1}^{2}(\omega)\!=\!\omega^{2}/s^{2}$ and $k_{i}^{2}(\omega)\!=\!k_{i}^{2}(0)\!+\!\beta_{i}\omega^{2}$ for $i\!\neq\!1$. Then we search for $\omega^{2}$ in the form of series by $k_{\tau}$, and on substituting the expansions of all quantities by small $k_{\tau}$ and $\omega$ into the equation (\ref{RipplonDispersion}), obtain
\begin{equation}
    \label{SmallOmega}
    \omega^{2}=\frac{\sigma}{\rho_{0}}\cdot k_{\tau}^{3}
    \left\{1\!-\!\frac{\sigma}{2\rho_{0}s^{2}} k_{\tau}\!+\!\frac{\sigma^2}{8\rho_{0}^{2}s^{4}} k_{\tau}^2\!+\!
        \frac{\sigma}{\rho_{0}}\frac{\beta}{i}
	\Big(k_{\tau}^2\!-\!\frac{\sigma}{\rho_{0}s^{2}} k_{\tau}^3\Big)\!+
	\!O\big(k_{\tau}^4\big)\right\},
\end{equation}
where $\beta\!=\!\sum_{i=2}^{S}\beta_{i}k_{i\,z}^{-1}(0,0)/2$.

The first two summands in the braces after unity are the first summands of the expansion of (\ref{compressibility}) and take into account compressibility. The next summands in (\ref{SmallOmega}), proportional to $\beta$, express the influence of roots $k_{i}$ with $i\!>\!1$, in particular of the roton ones. The latter give only small correction at small $\omega$, but with the increase of frequency they become of the order of the phonon summands and, as it will be shown below, they determine the asymptotic behavior of the curve $\omega(k_{\tau})$ in the proximity of $\omega\!=\!\Delta_{rot}$. The nonlinearity of dependence $k_{1}(\omega)$ itself, which determines whether the dispersion relation $\Omega(k)$ is normal or anomalous at small $k_{\tau}$,  gives to (\ref{RipplonDispersion}) the correction of the order $O(k_{\tau}^4)$.

Of particular interest is the behaviour of the dispersion curve close to $\omega\!=\!\Delta_{rot}$. Numerical solution of equation (\ref{RipplonDispersion}) shows, that the curve reaches the level of $\Delta_{rot}$ at some wave vector $k_{\tau}\!=\!k_c\!<\!k_{rot}$ (see Fig.~\ref{Fig1}). Let us introduce small parameters $\tilde\omega\!=\!\omega\!-\!\Delta_{rot}$ and $\tilde k\!=\!k_{\tau}\!-\!k_{c}$. In the region close to the minimum the dispersion of bulk excitations is well approximated by a parabola
\begin{equation}
    \Omega(k)\!\approx\!\Delta_{rot}\!+\!\frac{\hbar}{2\mu}(k\!-\!k_{rot})^{2}
\end{equation}
where $\mu$ is the ``roton mass". Then for the roton roots at $\omega\!\approx\!\Delta_{rot}$
\begin{equation}
    k_{2,3}\!\approx\!k_{rot}\pm\sqrt{\tilde\omega\cdot2\mu/\hbar}
\end{equation}
and
\begin{equation}
    \label{RotonRoots}
    k_{2,3\,z}\approx k_{rot\,z}
    \left(\mp1+\frac{k_{rot}}{k_{rot\,z}^{2}}\sqrt{2\frac{\mu}{\hbar}}\sqrt{\tilde\omega}\right),
\end{equation}
where $k_{rot\,z}\!=\!\sqrt{k_{rot}^{2}\!-\!k_{\tau}^2}\!>\!0$. Here the branch of square root is used, which gives $\sqrt{\tilde\omega}\!=\!|\sqrt{\tilde\omega}|$ for $\tilde\omega\!>\!0$ and $\sqrt{\tilde\omega}\!=\!i|\sqrt{\tilde\omega}|$ for $\tilde\omega\!<\!0$. In this way above $\Delta_{rot}$ the signs of $k_{2,3\,z}$ are such that $0<(-k_{2\,z})<k_{3\,z}$, and thus the wave packets comprised of these waves propagate away from the interface (in accordance with the selection rules used in \cite{PRB2008} and taking into account negative group velocity of $R^-$ rotons). Below $\Delta_{rot}$ the roots are defined by continuity and both give exponentially damped waves.

The non-linear dependence of $k_{2,3\,z}(\omega)$ near $\Delta_{rot}$ leads to the same square-root singularity in the equation (\ref{RipplonDispersion}):
\begin{equation}
    \label{RotonsEq}
    \frac{1}{i}\left(k_{2\,z}+k_{3\,z}\right)\approx b\sqrt{\tilde\omega},\quad\mbox{where}\quad
    b=-2i k_{rot}\sqrt{\frac{2\mu/\hbar}{k_{rot}^{2}\!-\!k_{c}^{2}}}=-i|b|.
\end{equation}
The expansions of other quantities in (\ref{RipplonDispersion}) in powers of $\tilde k$ and $\tilde\omega$ are trivial, and on substituting there (\ref{RotonsEq}), we obtain the asymptotic
\begin{equation}
    \label{RotonAsymptotic}
    \begin{array}[c]{l}
    \tilde k=-d\sqrt{\tilde\omega},\quad\mbox{where}\quad d=\frac{b}{a+c};\quad
    a=2\frac{\sigma}{\rho_0}\frac{\Delta_{rot}^2}{k_{c}^3};\\
    c=ik_{c}\left\{k_{1\,z}^{-1}(\Delta_{rot},k_c)+
	\sum\limits_{i=4}^{S}k_{i\,z}^{-1}(\Delta_{rot},k_c)-
		\sum\limits_{i=2}^{S}k_{i\,z}^{-1}(0,k_c)\right\}.
    \end{array}
\end{equation}
Here the branch of $\sqrt{\tilde\omega}$ is the same as before; quantities $a$ and $c$ are real.

Thus for $\omega\!=\!\Delta_{rot}\!-\!0$, we can rewrite the asymptote in original variables in the form
\begin{equation}
    \label{RotonAsympSimple}
    (k_{c}-k_{\tau})=\left|d\right|\cdot\sqrt{|\Delta_{rot}-\omega|}.
\end{equation}
We see from (\ref{RotonAsympSimple}) that the curve $\omega(k_{\tau})$ approaches the level $\omega\!=\!\Delta_{rot}$ at the top of an inverted parabola, and ends in the adhesion point, with zero derivative. There is no dispersion curve below $\Delta_{rot}$ with $k_{\tau}\!>\!k_{c}$. In \cite{Ripplon} the qualitative behaviour of the curve was derived, but the result was presented in an ambiguous form $\tilde\omega\sim{\tilde k}^{2}$. Even earlier the asymptote was obtained in theory in \cite{P&S95} as one of the possible variants, in a quite different approach, from general quantum-mechanic considerations. However, as opposed to that work, we have explicitly derived the coefficient $d$, which depends only on the bulk excitations spectrum $\Omega(k)$ and surface tension $\sigma$. The constant $k_{c}$, which is present in the coefficient as a parameter, is in turn obtained by numerical solution of algebraic equation (\ref{RipplonDispersion}) at $\omega\!=\!\Delta_{rot}$ (see next section).

It should be noted also, that close to $\omega\!=\!\Delta_{max}$ the roots $k_{1,2\,z}(\omega)$ have the same behaviour as $k_{2,3\,z}(\omega)$ in the considered case close to $\Delta_{rot}$. Therefore, if there was a common point of the ripplon dispersion curve with the line of the maxon level, the curve would also approach it with zero derivative and end at the adhesion point. However, the numerical solution (see below) shows, that such points do not actually exist.

Eq. (\ref{RotonAsymptotic}) gives also the asymptotic behavior of the curve above $\Delta_{rot}$. When $\tilde\omega\!>\!0$, $\tilde k$ is imaginary, which means that the surface solution dissolves on the distances of the order of $|\mathrm{Im} k_{\tau}|^{-1}\!=\!|\mathrm{Im}\tilde k|^{-1}$, decomposing into rotons. This is indicated by the structure that the solution (\ref{solution-waves}) takes in this case. At $\tilde\omega\!>\!0$ the two roton waves $\sim\!\exp(ik_{2,3\,z}z)$ are running waves in the $z$ direction. The correct choice of signs of $k_{2,3\,z}$ ensured that these waves carry energy away from the surface, and wave packets comprised of them propagate away from the surface. A different choice of signs would correspond to the processes of rotons reflection from the free surface, extensively studied in \cite{PRB2008}. The probabilities of either $R^-$ or $R^+$ roton's creation can be obtained by calculating the relative z-components of energy flows in the two corresponding waves.

Taking into account that energy is carried with group velocity, and average energy density in a wave with velocity amplitude $\mathbf{v}_{i}$ and group velocity $u_{i}$ is $u_{i}\rho_{0}|\mathbf{v}_{i}|^2$, we get that the $z$-component of energy flow in this wave is
\begin{equation}
    \label{flux}
    Q_{i}=\frac{|k_{i}k_{i\,z}|}{2\rho_{0}\omega^{2}}|u_{i}r_{i}^{2}|.
\end{equation}
As mentioned above, the roots $k_{i\,z}$ that lie in $\mathcal{C}_{+}$ are either real (i.e. which are shifted up from the real line; $k_{2,3\,z}$ are the only ones for $\omega\!>\!\Delta_{rot}$ and $k_{\tau}\!\approx\!k_{c}\!<\!k_{2}(\omega)$), or imaginary (as $k_{1\,z}$), or break up into pairs that are related as $k_{i\,z}\!=\!-k_{j\,z}^{\ast}$. When taking this into account and calculating (\ref{flux}), we obtain that
\begin{equation}
    \label{flux_simple}
    Q_{2,3\,z}\sim |k_{2,3\,z}|.
\end{equation}
So in the zero approximation by $\tilde\omega$ the probabilities of the surface excitation with energy greater than $\Delta_{rot}$ (or ``high-energy ripplon" for short) decaying into an $R^-$ or $R^+$ roton are equal to $1/2$. With the increase of frequency, decay into $R^+$ roton becomes more probable.

The derived probabilities should be applicable to a physical situation, when the free surface of superfluid helium is excited externally with characteristic frequency a little above $\Delta_{rot}$. Then high-energy ripplons should be created and consequently decay into rotons on distances of the order of $|\mathrm{Im} k_{\tau}|\!=\!|\mathrm{Im}\tilde k|^{-1}$ from the source.

Above $\Delta_{rot}$ and between the phonon and $R^-$ roton curves we have $k_{2\,z}$ and $k_{3\,z}$ real, and between the $R^-$ and $R^+$ roton curves only $k_{3\,z}$ is real. Evidently, in both cases, those real roots introduce nonzero imaginary parts into Eq. (\ref{RipplonDispersion}), and it has no solution in the plane of real $\omega$ and $k_{\tau}$. The imaginary part of $k_{\tau}$ should be of the order of the real part, which means that the excitations are unstable and quickly decompose into rotons.

\section{Numerical solution and new ripplon branch}

For numerical solution it is convenient to rewrite the equation (\ref{RipplonDispersion}). As mentioned above, the structure of the roots $k_{i\,z}$ is such, that only their imaginary parts contribute to (\ref{RipplonDispersion}), and for them $\mathrm{Im}k_{i\,z}\!>\!0$. Then we rewrite the sums including $i\!=\!1$, and taking into account that $k_{1\,z}(\omega\!=\!0)\!=\!ik_{\tau}$, we can represent Eq. (\ref{RipplonDispersion}) in the form
\begin{equation}
    \label{RipplonDispersionNum}
    \omega^2=\frac{\sigma}{\rho_{0}}\cdot k_{\tau}^2
    \left\{k_{\tau}+
    \sum\limits_{i=1}^{S}
            \Big|\mathrm{Im}\sqrt{k_{i}^{2}(\omega)\!-\!k_{\tau}^{2}}\,\Big|-
    \sum\limits_{i=1}^{S}
            \Big|\mathrm{Im}\sqrt{k_{i}^{2}(0)\!-\!k_{\tau}^{2}}\,\Big|
    \right\}.
\end{equation}
These sums taken over all $k_{i\,z}$ in $\mathcal{C}_{+}$ are now the same as the sums over all the $S$ roots $k_{i}^2$ of equation $\Omega^{2}(k^2)\!=\!\omega^{2}$. There is no need to sort the roots and define whether each lies in $C_{+}$ or not. The functions $k_{i}^{2}(\omega)$ are obtained numerically as all the roots of the corresponding polynomial equation. The equation (\ref{RipplonDispersionNum}) is valid for the dispersion relation of ripplons in the regions where they are stable, i.e. where all $k_{i\,z}$ are not real. Its numerical solution in the range of $0\!\leq\! k_{\tau}\!\leq\!3${\AA}${}^{-1}$ gives the curves shown on Fig.~\ref{Fig1}.

We see that, as predicted, the computed curve at small wave vectors is close to the classical $k_{\tau}^{3/2}$ law, but deviates from it at larger $k_{\tau}$ and approaches the level of $\Delta_{rot}$ at the top of an inverted parabola at $k_{c}\!=\!1.27${\AA}${}^{-1}$. As predicted in previous work \cite{Ripplon}, the use of better approximation of the bulk dispersion relation stretched the curve in this region towards the roton minimum $k_{rot}$, if compared with the results of \cite{Ripplon}. The reason for this is that the rough approximation used in \cite{Ripplon} underestimated the ``roton mass" by factor of order of $1.5$, so that the curvature radius of the dispersion curve at the roton minimum was overestimated. As the ripplon dispersion curve near $\Delta_{rot}$ is determined by the asymptotes of the roton roots there, the rough approximation overestimated also the curvature radius of the ripplon dispersion curve at the adhesion point.

There is no ripplon solution in the region on the high $k$ side of the $R^+$ roton dispersion curve (see Fig.~\ref{Fig1}), for $\omega\!\in\!(\Delta_{rot},\Delta_{max})$. If we rewrite the equation (\ref{RipplonDispersion}) in the form $F(\omega,k_{\tau})\!=\!0$, then in this area the main contribution to $F(\omega,k_{\tau})$ can be shown to be provided by the summand $\sim k_{2\,z}$, which thus prevents it from turning to zero. However, if we further increase $\omega$ while moving along the curve $k_{3}(\omega)$, as we get close to the region where $\Omega(k)/k$ reaches its maximum, the curve turns down and aims to the point of instability \cite{EndSpectrum} in almost straight line, the left hand part of Eq. (\ref{RipplonDispersion}) starts to change slowly. At the same time, in this region the structure of the roots $k_{i}^{2}(\omega)$ is qualitatively the same as in the neighbourhood of an inflection point of the curve $\Omega^{2}(k)$. This leads to rapid changes of some of the roots with $i\!>\!3$ (as near the inflection point two of the complex-conjugate roots tend to the same real value with the usual square-root asymptotic), which give significant contribution to $F(\omega,k_{\tau})$.Thus  here the solution is determined by the right-hand part of Eq. (\ref{RipplonDispersion}) and the functions $k_{i\,z}$.

Therefore, when we search for the solutions on the $R^+$ roton curve, two are found above $\Delta_{max}$. The first point is at $2.52(2)${\AA}${}^{-1}$ and $16.65(5)$K, almost exactly at the maximum of $\Omega(k)/k$, and the second is at $2.75(5)${\AA}${}^{-1}$ and $17.3(1)$K, close to the point of instability $\omega\!=\!2\Delta_{rot}$ (estimation of errors is made by comparison of the results given by different approximation polynomials, see appendix). The ripplon dispersion curve $\omega(k_{\tau})$ between them sticks closely to the bulk dispersion from below, and their two common points are adhesion points. This is partly the reason for the deviation between the curves being extremely small. At $k_{\tau}\!=\!2.6${\AA}${}^{-1}$, midway between the two end points, the deviation is $1.6\cdot10^{-3}\,K$ or $0.7\cdot10^{-3}${\AA}${}^{-1}$, which is too small to see on the scale of the main graph. So in the inset to Fig.~\ref{Fig1} we show this region expanded, and the separation of the  ripplon dispersion curve from the roton dispersion curve is greatly exaggerated.

Let us show that the common points of the ripplon dispersion curve $\omega(k_{\tau})$ and the $R^+$ roton dispersion curve $k_{3}(\omega)$ should indeed be adhesion points. There is the summand $k_{3\,z}$ in $F(\omega,k_{\tau})$, which turns to zero as $\sqrt{k_{\tau}\!-\!k_{3}(\omega)}$ on the $R^+$ roton branch. So the gradient of $F(\omega,k_{\tau})$ on the plane $(\omega,k_{\tau})$ tends to infinity on the curve $k_{\tau}\!=\!k_{3}(\omega)$ and is directed normal to the curve. From the other side, the gradient of $F(\omega,k_{\tau})$ is directed normal to the curve $\omega(k_{\tau})$, which is its level curve $F=0$. Therefore in the common points of the two curves the angle between them is equal to zero, and those points are adhesion points.

We see now, that in case there are common points of the curve $\omega(k_{\tau})$ with the boundaries of its possible existence (i.e. curves $k_{1}(\omega)$ and $k_{3}(\omega)$) or the lines $\omega\!=\!\Delta_{rot,max}$ of extremums of the bulk dispersion $\Omega(k)$, those points can only be adhesion points (with the exception of point $k_{\tau}\!=\!0$). Numerical solution shows that there are three such points altogether -- two on $k_{3}(\omega)$ and one at $\omega\!=\!\Delta_{rot}$.

\begin{figure}[!ht]
\begin{center}
\includegraphics[viewport=30 401 489 698, width=0.75\textwidth]{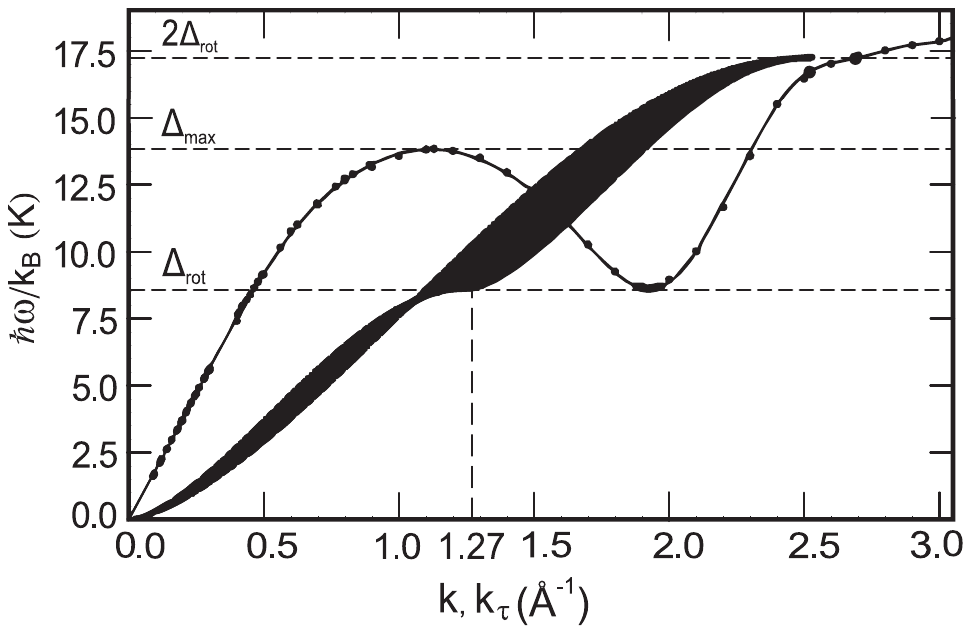}
\parbox{0.9\textwidth}{\caption{\label{Fig2} The set $A$ (black region) consists of the points, which correspond to excitations which can decay into two collinear ripplons with energy $<\Delta_{rot}$, from energy-momentum conservation considerations. As the new ripplon branch lies to the right of $A$, it is stable with regard to this process.}}

\end{center}
\end{figure}

At $k_{\tau}\!=\!2.6${\AA}${}^{-1}$ the value of $k_{3\,z}$ is close to its maximum $0.06${\AA}${}^{-1}$ on this branch of the ripplon curve, and the penetration depth of the ripplon solution, also determined by the $\sim\!k_{3\,z}$ summand here, to its minimum $\delta\sim |k_{3\,z}|^{-1}\sim 16${\AA}, tending to infinity at the end points. The deviation between the two curves is of the second order by the small parameter $k_{3\,z}$, as $|k_{3\,z}|\sim\sqrt{k_{\tau}\!-\!k_{3}}$.

The relatively large penetration depth of the solution means that macroscopic films or the surface of bulk helium are needed to observe these ripplons. They will not be seen on films of a few mononlayers. The high-energy ripplons should exist and be observable on saturated films of He~II \cite{SaturatedFilms}, which have a typical thicknesses of $300${\AA}. In the same way the penetration depth of the ripplons close to the roton gap (see Fig.~\ref{Fig1}) tends to infinity when $\omega$ tends to $\Delta_{rot}$. So the ending point of the dispersion curve $(\Delta_{rot},k_{c})$ can also be observed only in thick enough films of He~II.  Thus in the two most interesting regions of their dispersion curve, the ripplon solutions have large penetration depths, much larger than the characteristic distances, of several monolayers thickness, of the changes of the density profile at the free surface.

The numerical calculations were carried out for different approximation polynomials $P_{S}$ of function $\Omega^{2}(k^{2})$, with $S\!=\!18$ and $S\!=\!21$. Also different polynomials of power $21$ were obtained by approximating the experimental data supplemented by different subsets of the points, extracted from the spline (see appendix). The high-energy ripplons solution exists in all cases, and for different $P_{S}$ its ending points differ insubstantially. The deviation between the curves $k_{3}(\omega)$ and $\omega(k_{\tau})$ remains very small, much less than the deviation between the approximation curves themselves.

Stability of the new ripplon branch with regard to decay into two ripplons with energies less than $\Delta_{rot}$ can be checked graphically. In the process the energy and momentum should be conserved, so if we denote the initial high-energy ripplon by index $0$ and the resulting two quasiparticles by $1$ and $2$, we have
\begin{equation}
	\label{stab-om}
	\omega_{0}=\omega_{1}+\omega_{2}\quad\mbox{and}\quad
	\mathbf{k}_{0}=\mathbf{k}_{1}+\mathbf{k}_{2}.
\end{equation}
Then $k_{0}<k_{1}+k_{2}$. If we define the set $A$ of points $(\omega,k_{\tau})$ that obey the conditions $\omega=\omega_{1}+\omega_{2}$ and $k_{\tau}=k_{1}+k_{2}$ for all $(\omega_{1},k_{1})$ and $(\omega_{2},k_{2})$ satisfying the ripplon dispersion relation below $\Delta_{rot}$, then the point $(\omega_{0},k_{0})$ should be situated to the left of $A$, in order for the process of its decay into two ripplons to be allowed. This set is shown on Fig.~\ref{Fig2}, and we see that $A$ is to the left of the new ripplon branch, thus it is stable with regard to decay into two ripplons below $\Delta_{rot}$. In a similar way it can be shown that it is stable with regard to decay into three and more ripplons below $\Delta_{rot}$.

It is also natural to check for the stability of the solution with regard to changes in surface tension $\sigma$. While in this work we assume, for simplicity, that $\sigma$ is constant ($\sigma\!=\!3.544\,N/m$ at zero temperature \cite{neutron}), it is known (see \cite{SurfaceTension}) that better agreement with experiment is achieved if its curvature dependence is taken into account. For small $\omega$ this should only change the numerical results insignificantly. However, there is no data yet for surface tension at energies $\geq15\,K$, which correspond to the end of bulk spectrum, so the only thing we can do is to solve the equation with different values of $\sigma$ and see if the very high energy ripplons still exist. It appears, that if we decrease $\sigma$, the end points of the ripplon dispersion curve slide apart along $k_{3}(\omega)$ and its length increases. At the value $\sigma\!=\!1.75\,N/m$, the lower adhesion point reaches $16\,K$. If we, on the contrary, increase $\sigma$, then the end points slide towards each other and the solution disappears when $\sigma$ reaches the value of $8.7\,N/m$ ($2.5$ times value at $T\!=\!0$). It should be noted, however, that any realistic dependence $\sigma(k_{\tau})$ can be inserted in Eq. (\ref{RipplonDispersion}) without complicating the numerical solution.

It should be noted, that as the initial equation for pressure (\ref{EQP}) was first obtained in order to describe the superfluid-solid interface \cite{PRB2008}, the interface was assumed to be sharp, so this relatively simple generalization of nonlocal wave equation to half-space could be used, with the modified area of integration but the same kernel as in the infinite space. The same assumption of interface sharpness was maintained in this work for the calculation of ripplon's dispersion relation. In reality the density profile of the free surface of superfluid helium is rather smooth, and there is a transition layer of several monolayers thickness. However, when we consider ripplons with large penetration depths --- either those close to $\Delta_{rot}$ or the ripplons of the new branch, close to $k_{3}(\omega)$ --- or slow rotons close to the roton minimum, the major part of the energy of the wave is stored outside of the transition layer. Therefore the latter can be neglected, and the surface can be considered sharp (i.e. the kernel h(r) in Eq. (1) is the same as in the bulk fluid). This is also justified by the fact that the obtained results are in rather good agreement with both experiment and the results of calculations based on density-functional approach \cite{P&S95}, in which the smooth transition from liquid to vapour is reproduced.

\section{Conclusions}
In this work we have investigated the dispersion relation of the surface excitations of superfluid helium, ripplons. in the dispersive hydrodynamics approach, in which the only input ``parameter" is the dispersion relation of the bulk medium. We use the approximation of the dispersion relation by a polynomial and derive the equation for the ripplon's dispersion in terms of the roots of the corresponding polynomial equation. In this work, as opposed to \cite{Ripplon}, we use the polynomial of arbitrary power $S$, which can approximate the experimentally measured curve with any given accuracy in the full range of wave vectors, from zero to the point of instability.

The ripplon dispersion curve $\omega(k_{\tau})$ is investigated analytically both at small frequencies $\omega\!\rightarrow0$ and close to the level of roton minimum $\Delta_{rot}$, and the expansions are derived. It is shown to deviate slightly from the classic $k_{\tau}^{3/2}$ law at small $k_{\tau}$, but the deviation increases with greater $k_{\tau}$ and it approaches the level of $\Delta_{rot}$ at the top of an inverted parabola branch at $k_{\tau}\!=\!1.27${\AA}${}^{-1}$.

The probabilities of the decay of unstable ripplons above $\Delta_{rot}$ into $R^-$ and $R^+$ rotons are calculated and in zero approximation by $(\omega-\Delta_{rot})$ are shown to be equal to $1/2$.

The equation is solved numerically for $S\!=\!18$ and $S\!=\!21$. Besides the ripplon dispersion curve below $\Delta_{rot}$, a high-energy solution is found above the maxon level $\Delta_{max}$. It follows closely the bulk dispersion curve from $2.5${\AA}${}^{-1}$ to $2.7${\AA}${}^{-1}$ on distances $\sim\!10^{-3}\,K$ and its end points on the bulk dispersion curve are shown to be adhesion points. This solution is shown to be stable with regard to variation of approximation polynomials and to changes of surface tension in wide range. It is also stable with regard to decay into two or more ripplons with energies $<\!\Delta_{rot}$.

We hope that evidence for existence of the unusual ripplon branch above the roton gap, found in this work, stimulates new experiments, that would confirm the predictions of the theory.

\acknowledgements{We are grateful to EPSRC of the UK (grant EP/F 019157/1) for support of this work.}

\section*{Appendix. Fitting the dispersion curve of He~II}

In order to solve numerically Eq. (\ref{RipplonDispersion}), we have to prepare the approximating polynomial $\Omega^{2}(k^2)$. We use the following scheme for that. First, we construct least-square roots approximation of experimental data \cite{neutron} for $\Omega^{2}/k^{2}$, as function of $k^2$, by a polynomial of high enough power:
\begin{equation}
    \label{OmegaAppr}
    \frac{\Omega^{2}(k)}{k^{2}}=M_{S-1}(k^{2})=a_{0}+a_{1}k^{2}+\ldots+a_{S-1}k^{2(S-1)}.
\end{equation}
This way $\Omega^{2}\!\sim\!k^2$ at small $k$ regardless of approximation, which is important. The polynomials of powers 17 and 20 provide curves close enough, so we made the calculations for $M_{S-1}$ polynomials with $S\!=\!18$ and $S\!=\!21$. In order to suppress large amplitude high-frequency oscillations at high $k^2$, which occur because of sparseness of experimental points in the maxon-roton region (except for around the roton minimum), we supplement the data in this region by additional points taken from its spline (using different subsets of the points from the spline we obtain slightly different approximation polynomials $M$; the results of the calculations for the ripplon's dispersion for different  $M$ can be shown to differ insubstantially).

The resulting polynomial gives good approximation of the data in the given interval of $k$, which we choose as $k\!\in\![0,3]${\AA}${}^{-1}$, but at higher $k$ it has large amplitude oscillations, is nonmonotonic and turns into zero. The polynomial has to be further corrected in our case, to make it monotonic at $k\!>\!3${\AA}${}^{-1}$. Otherwise additional real roots of equation $\Omega^{2}(k)\!=\!\omega^2$ for the considered $\omega$ would appear, which would lead to existence of running waves in solution (\ref{solution-waves}) at $k_{\tau}\!<\!3${\AA}${}^{-1}$, which should be a surface solution. The additional zeros of $\Omega^{2}(k)$ for real $k$ also should not exist for the solution given in \cite{PhNT} to be valid. In order to correct the polynomial, we increase its highest power coefficient $a_{S-1}$ until the oscillations vanish and beyond the approximation interval $M$ becomes monotonic. The high power $S$ ensures that, while beyond $3${\AA}${}^{-1}$ function $\Omega^{2}(k)$ now rises as $(k^{2})^{S}$, the changes to it at $k\!<\!3${\AA}${}^{-1}$ are negligible. The resulting polynomial approximates experimental data on the chosen interval $k\!\in\![0,3]${\AA}${}^{-1}$ and obeys all the necessary additional conditions. If $a_{S-1}$ is negative, the trick does not work, and we just have to add the summand of even higher power with the coefficient that eliminates the oscillations at $k\!>\!3${\AA}${}^{-1}$, but is small enough for the summand to be negligible at $k\!<\!3${\AA}${}^{-1}$. The scheme can be used with a polynomial of any power, which can in principle approximate the experimental data with any given precision. One of the used sets of coefficients for $S\!=\!21$ is presented in table \ref{Table}.
\begin{table}
\begin{tabular}{||c|c||c|c||c|c||}\hline\hline
$\quad a_{0}\quad $&334.38          &   $\quad a_{7}\quad $&1536.6172803      & 
	$\quad a_{14}\quad$ &$2.0431806958\cdot10^{-4}$   \\
$a_{1}$&441.78          &  $a_{8}$&$-504.57796831$     &
       		$a_{15}$&$-1.25670939443\cdot10^{-4}$ \\
$a_{2}$&$-2680.24$    & $a_{9}$&124.559633213   &
        $a_{16}$&$1.24161912937\cdot10^{-5}$\\
$a_{3}$&5515.4755    &   $a_{10}$&$-23.052039054$    &     
	 $a_{17}$&$-6.988403321\cdot10^{-7}$\\
$a_{4}$&$-6881.11663$  &  $a_{11}$&3.13661307342    & 
	$a_{18}$&$2.434762814\cdot10^{-8}$\\
$a_{5}$&5809.52885   &   $a_{12}$&$-0.296185514695$    &     
	 $a_{19}$&$-4.918385291\cdot10^{-10}$\\
$a_{6}$&$-3490.147348$   &   $a_{13}$&0.0157637822729    & 
	 $a_{20}$&$4.43529347\cdot10^{-12}$\\ \hline\hline
\end{tabular}\\[0.3cm]
\caption{\label{Table} Coefficients for $S=21$}
\end{table}

The number of significant digits is large because of high powers of argument involved in Eq. (\ref{OmegaAppr}). The resulting ripplon dispersion curves, obtained when using different approximations, are indistinguishable below $\Delta_{rot}$ and the same branch manifests itself above $\Delta_{max}$. This serves as the indication of the stability of Eq. (\ref{RipplonDispersion}) for the ripplons' dispersion with regard to changes in the approximation polynomial. 

The actual calculations were carried out with precision greater than given in the table, but the presented coefficients are accurate enough to reproduce the dispersion of bulk excitation up to $3${\AA}${}^{-1}$.


\begin{thebibliography}{9}

\bibitem{KingWyatt} P.\,J. King and A.\,F.\,G.\,Wyatt, \textit{Proc. Roy. Soc. Lond. A}, \textbf{322}, 355 (1971).

\bibitem{Rochesurften} P. Roche, G. Delville, N.\,J Appleyard and F.\,I.\,B. Williams, \textit{J. Low Temp. Phys.} \textbf{106}, 565 (1997).

\bibitem{Atkins-ripplons} K.\,R. Atkins, \textit{Can. J. Phys.} \textbf{31}, 1165 (1953).

\bibitem{exp} H.J.~Lauter, H.~Godfrin, V.L.P.~Frank and P.~Leiderer, \textit{Phys. Rev. Lett.}, \textbf{68}, 2484 (1992).

\bibitem{exp2} H.J.~Lauter, H.~Godfrin and P.~Leiderer, \textit{J. Low Temp. Phys.}, \textbf{87}, N 3-4, 425 (1992).

\bibitem{GrimesAdams} C.\,C. Grimes and G. Adams, \textit{Phys. Rev. Lett.} \textbf{36}, 145 (1976).

\bibitem{DeboraCoimbra} D. Coimbra, S.\,S. Sokolov, J.-P. Rino and N. Studart, \textit{Phys. Rev. B} \textbf{74}, 035411 (2006).

\bibitem{brownwyatt} M. Brown and A.\,F.\,G. Wyatt, \textit{J. Phys. Condens. Mat} \textbf{15}, 4717 (2003).

\bibitem{NayakEdwards} V.\,U. Nayak, D.\,O. Edwards and N. Masuhara, \textit{Phys. Rev. Lett.} \textbf{50}, 990 (1983).

\bibitem{wyatttuckercregan} A.\,F.\,G. Wyatt, M.\,A.\,H. Tucker and R.\,F. Cregan, \textit{Phys. Rev. Lett.} \textbf{74}, 5236 (1995).

\bibitem{volovik} G.\,E. Volovik, \textit{J. Low Temp. Phys.} \textbf{145}, 337 (2006).

\bibitem{P&S92} L.~Pitaevskii and S.~Stringari, \textit{Phys. Rev. B} \textbf{45}, 13133 (1992).

\bibitem{P&S95} A.~Lastri, F.~Dalfovo, L.~Pitaevskii and S.~Stringari, \textit{J. Low Temp. Phys.} \textbf{98}, Nos. 3/4, 227 (1995).

\bibitem{PRB} I.\,N. Adamenko, K.\,E. Nemchenko and I.\,V. Tanatarov, \textit{Phys. Rev. B} \textbf{67}, 104513 (2003).

\bibitem{PRB2008} I.\,N.~Adamenko, K.\,E.~Nemchenko and I.\,V.~Tanatarov, \textit{Phys. Rev. B} \textbf{77}, 174510 (2008).

\bibitem{PhNT} I.N.~Adamenko, K.E.~Nemchenko and I.V.~Tanatarov, \textit{Fiz. Nizk. Temp.} \textbf{32}, No.3, 255 (2006).

\bibitem{JLTP2006} I.N.~Adamenko, K.E.~Nemchenko and I.V.~Tanatarov, \textit{Journal of Low Temp. Phys.} \textbf{144}, No. 1-3, 13 (2006).

\bibitem{Atkins-hydro} K.R.~Atkins, \textit{Phys. Rev.} \textbf{116}, 1339 (1959).

\bibitem{Natsik} V. D. Natsik, \textit{Low Temp. Phys.} \textbf{33}, 999 (2007),
[\textit{Fiz. Nizk. Temp.}, \textbf{33}, 1319 (2007) (in Russian)]

\bibitem{Ripplon} I.N.~Adamenko, K.E.~Nemchenko and I.V.~Tanatarov, \textit{J. Phys.: Conf. Series} \textbf{150}, No.3, 032107 (2009).

\bibitem{EndSpectrum} L.~Pitaevskii, \textit{Sov. Phys. JETP} \textbf{36}, 1168 (1959).

\bibitem{neutron} R.J.~Donnelly, J.A.~Donnelly and R.N.~Hills, \textit{J. Low Temp. Phys.} \textbf{44}, 471 (1981).

\bibitem{L&L}  L.D.~Landau, E.M.~Lifshitz, \textit{Fluid Mechanics (Course of Theoretical Physics, Vol. 6)}, Pergamon Press, London (1987).

\bibitem {SaturatedFilms} L.C.~Jackson, L.G.~Grimes, \textit{Advances in Physics} \textbf{7}, Iss.28, 435 (1958).

\bibitem {SurfaceTension} M.~Iino, M.~Suzuki, and A.J.~Ikushima, \textit{J. Low Temp. Phys.} \textbf{61}, 155 (1985).


\end{thebibliography}
\end{document}